\documentclass[aps,prl,notitlepage,twocolumn,superscriptaddress,longbibliography,nofootinbib]{revtex4-1}
\pdfoutput=1

\usepackage[T1]{fontenc} 
\usepackage{lmodern}
\usepackage{amsmath,amsthm,amssymb,amsfonts}
\usepackage{scalerel}
\usepackage{mathtools}
\usepackage{graphicx}
\usepackage{subfigure}
\usepackage[normalem]{ulem}
\usepackage[colorlinks = true,
            linkcolor = blue,
            urlcolor  = blue,
            citecolor = blue,
            anchorcolor = blue]{hyperref}
\usepackage{mathrsfs}
\usepackage{bbold}
\usepackage{units}
\allowdisplaybreaks
\usepackage{bm}
\usepackage{braket}
\usepackage{makecell}
\usepackage[nameinlink]{cleveref} 
\usepackage{upgreek}
\usepackage{blindtext}
\usepackage{verbatim}
\usepackage{algorithm}
\usepackage[noend]{algpseudocode}
\makeatother

\usepackage[dvipsnames]{xcolor}
\usepackage{bbm}
\usepackage[bb=boondox]{mathalfa}
\usepackage{array}
\usepackage{multirow}
\usepackage{tabularx}
\usepackage{float}
\usepackage{dcolumn}
\usepackage{slashed}
\usepackage{braket}
\usepackage{verbatim}
\usepackage{multirow}
\usepackage{resizegather}
\usepackage{titlesec}
\usepackage[toc,page]{appendix}
\usepackage[export]{adjustbox}
\newcommand{\appendixhead}%
{\centering\textbf{\huge Appendices}
\vspace{0.25in}}

\definecolor{teal}{RGB}{0, 158, 115} 
\definecolor{morange}{RGB}{255, 127, 0}

\newcommand{\n}{\mathcal{N}\,}
\newcommand{\dd}{\mathcal{D}}

\DeclareMathOperator{\Tr}{Tr}

\begin{document}

\title{How Random Are Ergodic Eigenstates of the Ultrametric Random Matrices and the Quantum Sun Model?}

\author{Tanay Pathak\,\,\href{https://orcid.org/0000-0003-0419-2583}
{\includegraphics[scale=0.05]{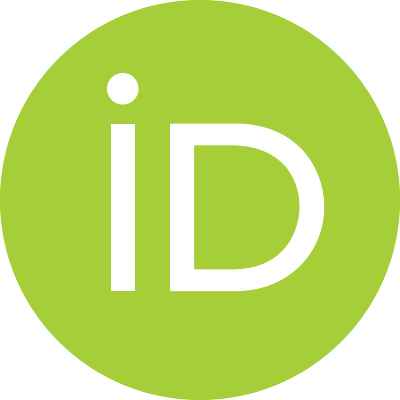}}}
\email{pathak.tanay@yukawa.kyoto-u.ac.jp}
\affiliation{Center for Gravitational Physics and Quantum Information, Yukawa Institute for Theoretical Physics,\\ Kyoto University, Kitashirakawa Oiwakecho, Sakyo-ku, Kyoto 606-8502, Japan}

\begin{abstract}
We numerically study the extreme-value statistics of the Schmidt eigenvalues of reduced density matrices obtained from the ergodic eigenstates. We start by exploring the extreme value statistics of the ultrametric random matrices and then the related Quantum Sun Model, which is also a toy model of avalanche theory. It is expected that these ergodic eigenstates are purely random and thus possess random matrix theory-like features, and the corresponding eigenvalue density should follow the universal Marchenko-Pastur law. Nonetheless, we find deviations, specifically near the tail in both cases. Similarly, the distribution of maximum eigenvalue, after appropriate centering and scaling, should follow the Tracy-Widom distribution. However, our results show that, for both the ultrametric random matrix and the Quantum Sun model, it can be better described using the extreme value distribution. As the extreme value distribution is associated with uncorrelated or weakly correlated random variables, the results hence indicate that the Schmidt eigenvalues exhibit much weaker correlations compared to the strong correlations typically observed in Wishart matrices. Similar deviations are observed for the case of minimum Schmidt eigenvalues as well . Despite the spectral statistics, such as nearest neighbor spacing ratios, aligning with the random matrix theory predictions, our findings reveal that randomness is still not fully achieved. This suggests that deviations in extreme-value statistics offer a stringent test to probe the randomness of ergodic eigenstates and can provide deeper insights into the underlying structure and correlations in ergodic systems.
\end{abstract}
\maketitle
~~~~~~~~~~~~Report Number: YITP-25-07

\section{Introduction}

The idea of localisation originally introduced by Anderson \cite{anderson1958absence} has been studied in great details in the presence of interactions as well. This phenomenon is widely referred to as many-body localisation (MBL) is interesting because such systems break ergodicity and fail to thermalize, and thus falls outside the realm of equilibrium statistical mechanics. It has been an active area of research in the past decade and studied in detail both theoretically \cite{Basko_2006, Gornyi_2005,Oganesyan:2007wpd,PhysRevB.77.064426,Pal_2010,PhysRevB.91.081103,Imbrie_2016,Cao:2023xez} (see \cite{Nandkishore_2015,Alet_2018,Abanin_2019,sierant2024mbl} for brief reviews and various other references) as well as experimentally \cite{Schreiber_2015,Smith_2016,L_schen_2017,Rispoli_2019,Guo_2020,Gong_2021,PhysRevResearch.4.013148,Filho_2022,L_onard_2023}.

In this regards the theory of quantum avalanches has been proposed as a mechanism of thermalization in interacting systems without translation invariance, wherein small ergodic exist in an otherwise nonergodic region \cite{PhysRevB.95.155129}. The mechanism thus explains why such a nonergodic region, with a small coupling to an ergodic region, may eventually thermalise \cite{PhysRevLett.119.150602,De_Roeck_2017,Thiery_2018,PhysRevB.99.195145,Gopalakrishnan_2019,PhysRevB.99.205149,Crowley_2020,PhysRevB.106.184208,PhysRevB.106.L020202,Morningstar_2022}. 
Quantum sun model (QSM) \cite{PhysRevLett.129.060602,PhysRevResearch.6.023030,PhysRevB.109.L180201,PhysRevB.110.134206} serves as the only toy model in which the avalanche mechanism  has been widely studied due to its analytical and numerical tractability.

Quantum entanglement has been an important tool to study these localization phenomena. The eigenstates of many-body Hamiltonian exhibit volume law for the entanglement entropy in the ergodic regime and area law in the localized regime \cite{Bardarson_2012,Kj_ll_2014,PhysRevB.91.081103,Nandkishore_2015,Alet_2018,Bera_2016}. However, the entanglement entropy captures only a small part of the full spectrum. Thus at times it might be useful to consider the whole entanglement spectrum \cite{Li_2008} and it has been employed successfully in the studies of MBL as well \cite{Yang_2015,Geraedts_2016,Serbyn_2016,Pietracaprina_2017}. 

In this work we aim to use the information of the entanglement spectrum to study the properties of ergodic eigenstates of the QSM. The model exhibits a transition from a localized regime to ergodic regime when the coupling strength parameter in the model is varied. It is well known that the eigenstates and eigenvalues in the ergodic regime are random and the behaviour of these are well predicted by random matrix theory \cite{berry1977level,marklof2001berry,PhysRevLett.52.1,bohigas1984spectral,casati1980connection,berry1985semiclassical} ( also see \cite{elyutin1988quantum,Guhr_1998,kriecherbauer2001random,Borgonovi_2016,Livan_2018,eynard2018randommatrices} and references therein for a more exhaustive list). Of interest to us in this work is the  Berry's conjecture \cite{berry1977regular,giannoni1991chaos} which says that the eigenstates of a quantum chaotic Hamiltonians would be a superposition of plane waves with random phases and Gaussian random amplitude with fixed wavelength and this gives rise to the famous Thomas-Porter law for the intensity of the random eigenstate \cite{porter1965statistical, mehta2004random}. Among other notable such connection is eigenstate thermalization hypothesis (ETH) \cite{PhysRevA.43.2046,PhysRevE.50.888,rigol2008thermalization,deutsch2018eigenstate,D_Alessio_2016} which is an important way to characterize the ergodic eigenstates and relate it to quantum thermalization and randomness. In connection to QSM, random matrix theory(RMT) has been useful in the study of the localized to ergodic transition \cite{PhysRevLett.129.060602,PhysRevResearch.6.023030,PhysRevB.109.L180201,PhysRevB.110.134206} using measures such as the eigenvalue correlation measure such as the eigenvalue spacing ratio \cite{Atas2013distribution, Atas_2013}. Recently in \cite{PhysRevResearch.6.023030} certain illuminating relations between a class of random matrices called the Ultrametric (UM) random matrices \cite{fyodorov2009anderson,rushkin2011universal,von2018phase,bogomolny2011eigenfunction,M_ndez_Berm_dez_2012,Gutkin_2011,von2017renormalization, Bogomolny_2018,von_Soosten_2019} were studied in great detail and many properties of eigenstates of UM random matrices and QSM were studied using inverse participation ratio, their derivatives and the entanglement entropy of their eigenstates.

We continue the study to elucidate the properties of the ergodic eigenstates of the UM random matrices and the QSM. It is already known that the eigenstate component distribution of the UM random matrices follow the Generalized hyperbolic distribution (GHD) \cite{Bogomolny_2018}, which has also been conjectured as yet another universal distribution alongside Porter-Thomas distribution. The study however reveals that for the QSM the fit with GHD is unsatisfactory, showing signs of its deviations from the behaviour of UM random matrices. Furthermore, the ergodic eigenstates are expected to be truly random, hence the eigenvalue of the reduced density matrices ( constructed out of these ergodic eigenstates) are expected to follow the Wishart distribution \cite{mehta2004random,haake1991quantum,forrester2010log}. Specifically, the eigenvalue density distribution of these eigenvalues are supposed to follow the Marchenko-Pastur law \cite{mehta2004random,forrester2010log}. However, we observe deviation from this universal distribution for both of them. Since these eigenvalues are strongly correlated, thus the distribution of maximum eigenvalue should obey the Tracy-Widom distribution \cite{tracy1996orthogonal}. On the other hand when the eigenvalues are uncorrelated or weakly correlated then the corresponding distribution should belong to the class of extreme value distributions (EVD) due to the Fisher–Tippett–Gnedenko theorem \cite{fisher1928limiting,gumbel1958}. Our study the eigenvalues does follow the EVD and thus have much weaker correlation than expected for ergodic eigenstates. These test thus provide some of the stringent tests to probe of the randomness of the ergodic eigenstates and suggest that ergodicity breaking should not be merely associated with the breakdown of the spectral statistics. 

We outline the paper as follows. In section \ref{sec: umrmt} we start by introducing the UM  random matrices and then discussed their properties:  eigenvector distribution, the Schmidt eigenvalues of the reduced density matrices constructed using its ergodic eigenstates and their deviation from the Marchenko-Pastur law, extreme value statistics of Schmidt eigenvalues including both the maximum and the minimum eigenvalues statistics. We then extend this study for the quantum sun model in section \ref{sec:qsm} and discuss qualitative differences between them. Finally we discuss the summary of the results and outlook in section \ref{sec:summary}.

\section{Preliminaries}
To make the further discussions self-consistent we introduce the preliminary notions that we will be using in the paper. 
Let us begin by introducing the entanglement spectrum and some of its properties. Let $\mathcal{H} = \mathcal{H}_{1} \otimes \mathcal{H}_{2}$ be a bipartite Hilbert space with $\text{dim}( \mathcal{H}_{1}) = \dd_{1}$, $\text{dim}( \mathcal{H}_{2}) = \dd_{2}$ and $\text{dim}( \mathcal{H}) = \dd_{1}\dd_{2}$. Let $\Psi$ be a state belong to this Hilbert space then we have for the following Schmidt decomposition of this state with respect to the basis state $\ket{i}_{1} \in \mathcal{H}_{1}$ and $\ket{i}_{2} \in \mathcal{H}_{2}$
\begin{equation}
    \Psi = \sum_{i_{1}=1}^{\dd_{1}} \sum_{i_{2}=1}^{\dd_{2}} c_{i_{1}j_{1}} \ket{i_{1}} \otimes \ket{i_{2}} = \sum_{i'=1}^{\dd_1} \sqrt{\lambda_{i'}} \ket{i'}_{1}\ket{i'}_{2},
\end{equation}
where $\lambda_{i'}$ are called the Schmidt eigenvalues and $\ket{i'}_{1}$, $\ket{i'}_{2}$ are the normalized eigenvalues of matrix $C^{\dagger}C$. 
The reduced density matrix of the subsystems are given by $\rho_{1}= C C^{\dagger}$ and $\rho_{1}= ((C^{\dagger}C)^{T}$, where $C$ is the coefficient matrix whose elements are $C_{ij}$.

From the structure of reduced density matrix above it is easy to deduce that they are of Wishart type. Since the trace of the reduced density matrix is 1, hence strictly speaking they belong to the ensemble of trace restricted Wishart matrices. If $\mathcal{G}$ is a structureless random matrix with Gaussian real or complex entries (hence it is non-Hermitan), then the reduced density matrices are given by the ensemble of following matrices 
\begin{equation}
    W= \frac{\mathcal{G} \mathcal{G}^{\dagger}}{\Tr(\mathcal{G} \mathcal{G}^{\dagger})},
\end{equation}
where the denominator ensures that the trace of the matrices is 1. The joint probability distributions of the eigenvalues of Wishart matrices is exactly known \cite{mehta2004random}. The average density of the Schmidt eigenvalue follow the Marchenko-Pastur law \cite{mehta2004random}, which for the special case when $\dd_{1}= \dd_{2}= \dd$ is given as by 
\begin{equation}\label{eqn:mplaw}
    \rho(\tilde{x})= \frac{1}{2 \pi} \sqrt{\frac{4- \tilde{x}}{\tilde{x}}}, \quad 0\leq \tilde{x} \leq 4,
\end{equation}
where $\tilde{x}= \dd \lambda_{i}$ are the suitable rescaled Schmidt eigenvalues. It is worthwhile to note that Marchenko-Pastur law is a universal distribution for the ensemble of correlation matrices, independent of the exact distribution of matrix elements given it has a finite moments of sufficiently larger order \cite{tao2012random}.

The eigenvalues of the Wishart matrices are strongly correlated and the probability density of the maximum eigenvalue follow the Tracy-Widom distribution \cite{tracy1996orthogonal}. Another important result for the extreme values is given Fisher–Tippett–Gnedenko theorem \cite{fisher1928limiting,gumbel1958} (also see \cite{Majumdar_2020} for a nice review) which states that the for a set of random variables: $\{X_{1},X_{2},\cdots , X_{N}\}$, which are identical and independent random variable with probability density $p(x)$, then the probability density of the the maximum of $X_{i}$ after suitable centering and rescaling follow one of the three limiting distribution : Fisher-Tippet-Gumbel, Weibull and Fr\'echet depending on the tail of the density of $p(x)$ is power law or faster than any power law and whether it is bounded or unbounded. The probability of suitably centered and rescaled $x \equiv y$ is given as follows 

\begin{align}\label{eqn:evd}
P(y ; \tilde{\xi})& = 
 \begin{dcases}
        \exp \left(-(1-\tilde{\xi} y)^{1 / \tilde{\xi}}\right)(1-\tilde{\xi} y)^{(1 / \tilde{\xi}-1)} & \tilde{\xi} \neq 0 \\
        \exp (-\exp (-y)) \exp (-y)& \tilde{\xi}=0.
 \end{dcases}
\end{align}
with $\tilde{\xi}$ being the shape parameter and $\tilde{\xi}= 0$, $\tilde{\xi}< 0$ and $\tilde{\xi}> 0$ implies Gumbel, Fr\'echet and Weibull distribution respectively.

In the presence of correlations these universal distributions are still valid \cite{leadbetter1988extremal}. However, Tracy-Widom distribution differ from these universal distribution because they arise for the case of strong correlations(that are present in the eigenvalues of the random matrices), whereas the Fisher–Tippett–Gnedenko theorem is valid for weak correlations.   It is also worthwhile to note that the results for both maximum and minimum eigenvalue, beyond these asymptotic universal distribution have also been obtain in RMT \cite{majumdar2008exact, Majumdar2011,PhysRevLett.100.240501,Vivo_2011,Kumar_2017,Forrester_2019}.

\section{Physical Random Matrix Ensembles: Ultrametric Random Matrices}\label{sec: umrmt}

The usual random matrix ensembles have a \emph{unphysical} condition that all the states interact with each other with approximately the same strength. However, in physical situation this is not usually the case and requires slight modification to the usual Gaussian ensembles. One way to incorporate these physical situations is to consider random matrix ensembles for which the variance of the elements is different as we move away from the diagonal. Amongst many of such ensembles are Random banded matrices (RBM), Power law random banded matrices (PLRBM) and Ultra metric random matrices (UMRM). We  focus our attention primarily to Ultrametric random matrices. We follow the construction strategy of the Hamiltonian as outlined in \cite{PhysRevResearch.6.023030}. The model is constructed as a sum of block diagonal random matrices $H_{j}$ with $j= 0,1, \cdots, L$, where $j$ is proportional to ultametric distance $d$\footnote{Ultrametric distance function $d(p,q)$ is defined such that it follows the strong triangle inequality $d(p,q) \leq max\{d(p,r), d(r,q \}$\cite{fyodorov2009anderson}}. At each $j$, the matrix structure of $H_{j}$ consists of $2^{L-k}$ diagonal block of size $2^{\n+k} \times 2^{\n+k}$. Each random block, $G^{i}$ is independently sampled from a GOE distribution and we normalize each block $H_{k}^{i}$ as follows
\begin{equation}
    H_{k}^{i} = \frac{G^{i}}{\sqrt{2^{\n+k}+1}}, \quad i = 1, \cdots, 2^{L-k}.
\end{equation}
The Ultra metric random is then given by 
\begin{align}
    H = H_{0} + J \sum_{j=1}^{L}\alpha^{j} H_{j}, \quad \alpha \in [0,1].
\end{align}

The specific construction is well suited for its application and similarity with the quantum sun model \cite{PhysRevResearch.6.023030}. The UM random matrices exhibit transition at $\alpha_{c} =  \frac{1}{\sqrt{2}}$.

For the numerical tests, throughout the paper,  we fix the value of $\alpha =0.9$, taking motivation from previous results where for this particular value of $\alpha$ the model has achieved the required \emph{RMT behaviour}. It is this attainment of (truly random) RMT type behaviour  that is attributed to ergodicity and our goal would be to present some stringent test to further challenge it. 
For the case of UM random matrices all the numerical results are obtained by taking $\n= 1, L= 11$, such that $\n+ L= 12 $. The choice is motivated by the numerical results obtained in \cite{PhysRevResearch.6.023030}. However, we expect the results to hold for other values of $\n$. For the case of QSM we use the choice $\n= 5, L= 9$ such that $\n+ L= 12 $. The density matrix is constructed using 300 eigenstates near the middle of the spectrum. The reduced density matrix constructed for both the models has dimension $2^{6} \times 2^{6}$, obtained by tracing out half of the Hilbert space of dimension $2^{12} \times 2^{12}$.

\subsection{Eigenvector distribution}
As a first step we would start with the analysis of the eigenvector statistics of the UM random matrices.  Let the \emph{random} real eigenvectors be $\ket{\psi}$. In some orthogonal basis $\ket{j}$, the components of eigenvector $\ket{\psi}$ are $x_{j}= \braket{j|\psi}$, then the jpdf of these components are uniform on the sphere 
\begin{equation}
    P(x_{1}, \cdots, x_{\dd}) = C_{\dd} \delta\left(1- \sum_{k=1}^{\dd}x_{k}^{2}\right).
\end{equation}
The marginal distribution of $x$ can be easily calculated and found to be
\begin{align}
    P(x)&=  \frac{\Gamma(n/2)}{\sqrt{\pi} \Gamma((n-1)/2)} (1-x)^{(n-3)/2}, \nonumber \\
    &\xrightarrow[]{x \rightarrow \infty} \sqrt{\frac{\dd}{2\pi} }e^{-\dd x^{2}/2}.
\end{align}
Hence we find that the rescaled eigenvector component $\tilde{x}_{i} \equiv \sqrt{\mathcal{D}} x_{i}$ follow the standard normal distribution. The intensity $y_{i} \equiv \sqrt{\mathcal{D}} x_{i}^{2}$, gives the famous Porter-Thomas law \cite{porter1965statistical, mehta2004random} for the intensity 
\begin{equation}
    P(y) = \frac{1}{\sqrt{2 \pi y}} e^{-y/2}.
\end{equation}

\begin{figure}[htbp]
    \centering
    \includegraphics[width=\linewidth]{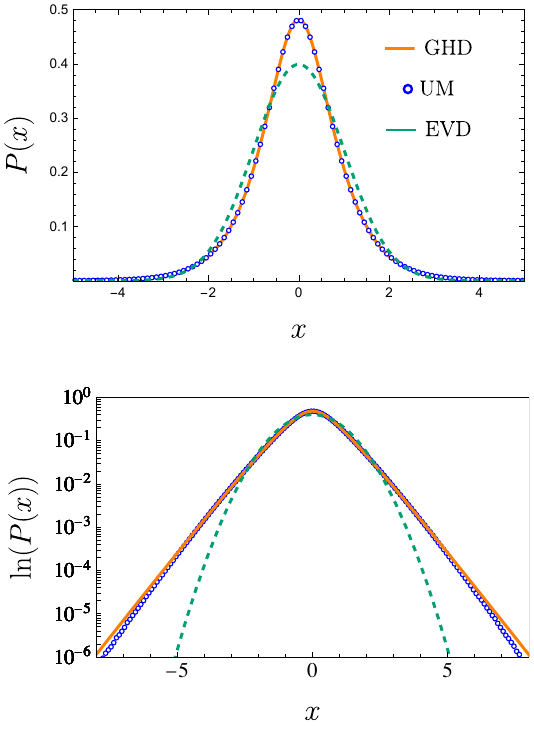}
    \caption{The top figure shows the distribution of rescaled eigenvector components,$x \equiv \sqrt{\dd} \psi$; $\dd= 2^{12}$, of UM random matrices with $\alpha= 0.90$. Solid red line are the fitted GHD distribution given by Eq.\eqref{eq:ghd} and \textcolor{teal}{teal} dashed lines denote the standard normal distribution. The bottom figure is the same as the top plot but with logarithmic $y-$ axis.}
    \label{fig:umghdfitall}
\end{figure}

Moving on the case of UM random matrices, taking the eigenvector components and rescaling them as $x \equiv \sqrt{\dd} \psi$; $\dd= 2^{\n+L}$, we can obtain the distribution as shown in Fig. \eqref{fig:umghdfitall} (blue markers) . We observe a clear deviation from the normal distribution, plotted as \textcolor{teal}{teal} dashed curve. Instead we observe a good fit with generalized hyperbolic distribution (GHD) as has been previously demonstrated in detail for both the power law random banded matrices and UM random matrices in \cite{Bogomolny_2018}. In the same it was also conjectured as a new universal distribution of extended states on equal footing with the Porter Thomas distribution. 
The PDF of the distribution is 
\begin{equation}\label{eq:ghd}
P (x)=  \frac{\sqrt{a } \left(c ^2+x^2\right)^{\frac{1}{2} \left(b -\frac{1}{2}\right)} K_{b -\frac{1}{2}}\left(a  \sqrt{x^2+c ^2}\right)}{\sqrt{2 \pi } c ^{b } K_{b }(a  b )},
\end{equation}
where $a, b$ and  $c$ are free parameters and $K_{n}(x)$ is the Bessel's $K$ function. The distribution can further be simplified to a two parameter distribution via the following substitution 
\begin{equation}
    a= \sqrt{\frac{\xi K_{b+1 }(\xi) }{K_{b}(\xi) }}, \quad c= \frac{\xi}{a}.
\end{equation}
The above substitution implies that the variance of the GHD is equal to 1 \cite{Bogomolny_2018}.
For the present case we found these free parameters to be : 

\begin{center}
$ b= 1.43068 , \quad \xi= 1.17378$.
\end{center}    
In Fig. \eqref{fig:umghdfitall} small deviations are observed for small $P(x)$, which might be due to slightly different fitting scheme in \cite{Bogomolny_2018} and ours. For our fitting we have considered all the eigenvector components where in the former only \emph{some} of the eigenvectors were considered. 

We would also like to note that similar deviations for the ergodic eigenstates, from the Porter-Thomas distribution, has been observed in \cite{pal2020}, with a similar study for the disordered Heisenberg model which is expected to host a many body localized phase \cite{PhysRevB.77.064426,Pal_2010,PhysRevB.91.081103,Cao:2023xez} ( see \cite{Alet_2018} and references therein). However, for that case a good fit was instead obtained using the exponential power distribution. 

\subsection{Deviations from Marchenko-Pastur distribution}\label{sec:ummp}

We now consider the Schmidt eigenvalues of the reduced density matrix which are constructed using the ergodic eigenstates. Since the eigenvalues corresponds to that of trace restricted Wishart matrices, in the large dimension limit of matrices we expect the Marchenko-Pastur (MP) distribution, which for the case when both the subsystem has equal dimensions $= \dd$, is given by 
\begin{equation}
    \rho(\tilde{x})= \frac{1}{2 \pi} \sqrt{\frac{4- \tilde{x}}{\tilde{x}}},
\end{equation}
where $\tilde{x} = \dd \lambda$ are the re-scaled eigenvalues.
\begin{figure}[ht]
    \centering
    \includegraphics[width=  \linewidth]{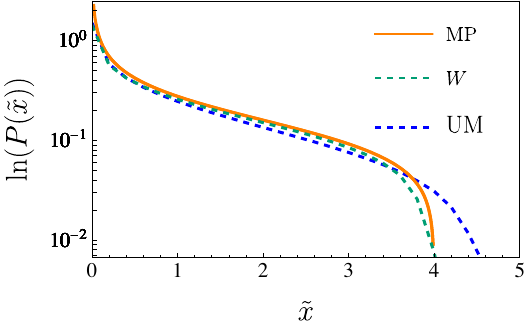}
    \caption{Distribution of rescaled eigenvalues, $\tilde{x}$ of the UM matrices shown as dashed blue curve. The orange solid line is the Marchenko-Pastur distribution and the dashed \textcolor{teal}{teal} line is the numerical curve obtained for the trace restricted Wishart matrices of size $2^{6} \times 2^{6}$, so as to match the size of reduced density matrix for the UM matrix case. We see a clear deviation from the Marchenko-Pastur law for the case of UM matrices, specially near the tail, while the Wishart matrices match quite well with the MP law for the same size matrices.}
    \label{fig:ummplaw}
\end{figure}
In Fig. \eqref{fig:ummplaw} we obtain the result for the UM matrices. The reduced density matrices are obtained using the eigenvectors of the UM matrices of size $2^{12} \times 2^{12}$ and tracing half the system of size $2^{6} \times 2^{6}$. The PDF of the eigenvalues of the resulting matrix are plotted along with MP law ( given by orange curve). We observe a clear deviation for the case of UM matrices (dashed \textcolor{blue}{blue}) near the tail of the distribution . For comparison we also obtain result for $2^{6} \times 2^{6}$ Wishart matrices, given by dashed \textcolor{teal}{teal} curve in Fig. \eqref{fig:ummplaw}. It is observed that although there is a mismatch at the end of the tail ( which can be attributed to finite size effect), the behaviour is same.
For further comparison, we compare the moment of MP distribution. The moments of the MP distribution can be easily evaluated and given by \cite{haake1991quantum}
\begin{equation}
    \mu_{k} = \frac{4^{k+\frac{1}{2}} B\left(k+\frac{1}{2},\frac{3}{2}\right)}{\pi },
\end{equation}
where $B(a,b)= \frac{\Gamma(a)\Gamma(b)}{\Gamma(a+b)}$. 

\begin{figure}[htbp]
    \centering
    \includegraphics[width=  \linewidth]{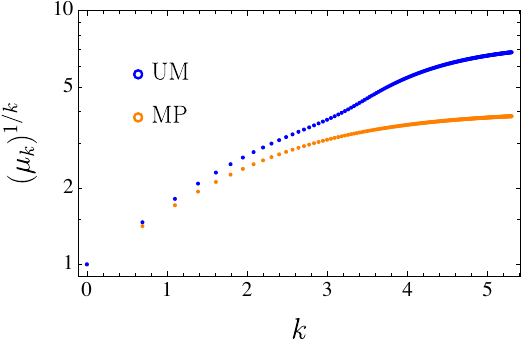}
    \caption{Comparison of $(\mu_{k})^{1/k}$; $\mu_{k}$ is the $k$-th moment, of the Schmidt eigenvalues of UM random matrices (blue markers) with the moments of Marchenko-Pastur distribution (orange markers).}
    \label{fig:ummom12}
\end{figure}

It is easy to see $\lim_{k \to \infty} (\mu_{k})^{1/k} = 4$. Hence we compare the $(\mu_{k})^{1/k}$ of MP distribution with the same obtained numerically for the UM random matrices in Fig.\eqref{fig:ummom12}. We observe that apart from the first moment none of the other moments match, showing a clear deviation from the MP law.  Such deviations from the MP law as observed for UM random matrices are thus interesting.

\subsection{Extreme value statistics of Schmidt eigenvalues}\label{sec:extum}
For the Wishart matrices the distribution of maximum eigenvalues after suitable centering and scaling follow the Tracy-Widom distribution \cite{tracy1996orthogonal}. For the case for reduced density matrix matrix we have $\rho =  \frac{W}{\Tr(W)}$, where $W= M M^{\dagger}$. The mean \cite{johnstone2001} of the distribution of $\lambda_{max}(W)$ is given as 
\begin{equation}
    \braket{\lambda_{max}(W)} \sim (\sqrt{\dd-1}+\sqrt{\dd})^{2},
\end{equation}
and the variance is given as 
\begin{equation}
    \sigma_{\lambda_{max}(W)} \sim (\sqrt{\dd-1} + \sqrt{n})\left( \frac{1}{\sqrt{\dd-1}} +  \frac{1}{\sqrt{\dd}}\right).
\end{equation}

\begin{figure}[ht]
    \centering
    \includegraphics[width=  \linewidth]{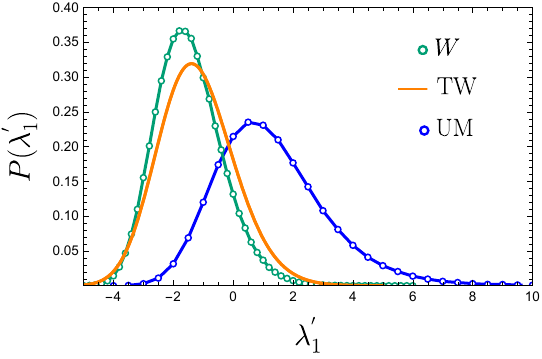}
    \caption{Distribution of maximum Schmidt eigenvalue, $\lambda'$, of the UM random matrices ( blue markers). The results are compared to the Tracy-Widom distribution ( blue solid line). The \textcolor{teal}{teal} markers are the results for the Wishart matrices of same dimension, $2^{6} \times 2^{6}$, for the comparison purposes. The blue and \textcolor{teal}{teal} line are guide to the eye.}
    \label{fig:umtw}
\end{figure}

Using this we obtain for reduced density matrix $\rho$
\begin{equation}
  \lambda_{max}(\rho) \sim \frac{4}{\dd}, \quad   \sigma_{\lambda_{max}(\rho)} \sim 2^{\frac{4}{3}} \dd^{-\frac{5}{3}}.
\end{equation}
We use these results to obtain the distribution of $\lambda_{max}$ for the UM random matrices. If $\lambda_{1}$ is the maximum Schmidt eigenvalue of the reduced density matrix, then we center and rescale it as
\begin{equation}
    \lambda^{'}_{1} = \frac{\lambda- \braket{\lambda_{max}(\rho)}}{\sigma_{\lambda_{max}(\rho)}}.
\end{equation}
We show the results in Fig. \eqref{fig:umtw}. It is clear from the numerical results that the distribution of $\lambda'_{1}$, $P(\lambda'_{1})$, for the UM matrices ( blue open markers) differ \emph{significantly} from the Tracy-Widom distribution $F_{1}$ ( solid orange curve). For comparison of the finite size effect due to small dimension of reduced density matrix, equal to $2^{6} \times 2^{6}$, we also obtain the results for the Wishart matrices of the same dimensions. It is observed that the results still deviate slightly from the Tracy-Widom distribution, which is due to the finite size effect as the Tracy-Widom distribution are expected to hold for large size matrices. To show this specific feature similar results are plotted for  Wishart matrices of dimension $2^{12} \times 2^{12}$ matrices, with $30,000$ realizations are shown in Fig. \eqref{fig:tww12} ( Appendix \eqref{app:tracywidom}). We observe that match is better for large matrices though still not very good.

Since we saw deviations from the well known Tracy-Widom distribution, it is imperative to understand the cause of it. Tracy-Widom distribution assumes that the eigenvalues are strongly correlated hence deviations from it should imply eigenvalues are weakly correlated. If this assertion is true then we expect the Fisher–Tippett–Gnedenko theorem \cite{fisher1928limiting,gumbel1958} to hold for the maximum eigenvalue of the reduced density matrix. To test this assertion we proceed same as before. We now redefine $\lambda_{1}$ as follows
\begin{equation}
    \tilde{\lambda}_{1} = \frac{\lambda_{1}-\tilde{\alpha}}{\tilde{\beta}},
\end{equation}
where $\alpha$ and $\beta$ are the free parameters that are to be determined. We then fit distribution of $\tilde{\lambda}_{1}$ with the 
extreme value distribution given by Eq. \eqref{eqn:evd}.

\begin{figure}[ht]
    \centering
    \includegraphics[width=  \linewidth]{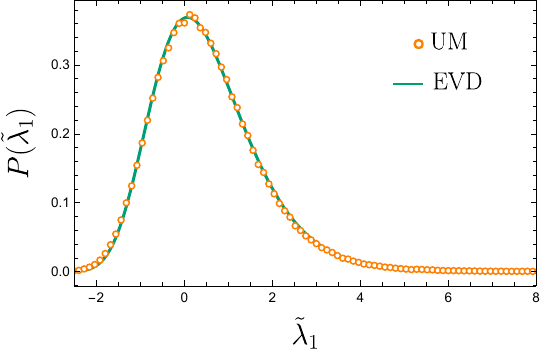}
    \caption{Distribution of rescaled eigenvalues. $\tilde{\lambda}_{1}$ for UM matrices shown as dashed blue curve. The \textcolor{teal}{teal} solid line is the fitted extreme value distribution and the orange markers are the numerical results obtained for the QSM.}
    \label{fig:um90evd}
\end{figure}
The obtained results for the UM random matrices are shown in Fig. \eqref{fig:um90evd}. We obtained a good match with the extreme value distribution given by Eq. \eqref{eqn:evd} with the following fitting parameters 
\begin{center}
    $\tilde{\alpha}= 6.410 \times 10^{-2},\tilde{\beta}= 3.822\times 10^{-3}$, $\tilde{\xi}= 7.246 \times 10^{-2}$,
\end{center}
which suggest \emph{Weibull} distribution. This confirms the assertion that the correlations are not Wishart like even for the ergodic eigenstates. This is in sharp contrast with the result obtained using nearest neighbour distribution captured using spacing ratios \cite{PhysRevResearch.6.023030}. The reason for this sharp contrast can be attributed to the fact that nearest neighbour distribution and usual spacing ratios captures \emph{only} the short range correlation while the presence of Tracy-Widom distribution would imply strong correlation which are not only short ranged but long ranged as well. It is exactly these higher order correlations whose information is captured by the maximum eigenvalue distribution. It would thus be useful to study higher order spacing ratios in present context \cite{PhysRevB.98.104305,Srivastava_2018,Buijsman_2019, xu2019,Rao_2022} and in general to study their utility in the study of ergodicity of eigenstates \footnote{It is important to note that in general,it suffices to consider the nearest neighbor interaction without dwelling much into the higher order correlations. However, it is important to note that the true RMT behaviour would imply correlations of higher orders as well.}. 

Another interesting case of \emph{strongly correlated} random variables where the minimum distribution can be analytically obtained is the case of minimum eigenvalues of reduced density matrix \cite{majumdar2008exact}, further generalized for unequal dimensions of sub-systems in \cite{chen2010smallest}. It was found that the minimum eigenvalue of the reduced density matrix corresponding to a real random eigenvector follow the following distribution
\begin{align}\label{eqn:lmin}
    P( \lambda_{min})= 
\frac{\lambda_{min}^{-\frac{\mathcal{D}}{2}} (1-\mathcal{D} \lambda_{min})^{\frac{1}{2} \left(\mathcal{D}^2+\mathcal{D}-4\right)} \, 
\left(\mathcal{D} \, \Gamma (\mathcal{D}) \, \Gamma \left(\frac{\mathcal{D}^2}{2}\right)\right) \, }
{2^{\mathcal{D}-1} \, \Gamma \left(\frac{\mathcal{D}}{2}\right) \, \Gamma \left(\frac{1}{2} \left(\mathcal{D}^2+\mathcal{D}-2\right)\right)} \nonumber \\
\times _2F_1\left(\frac{\mathcal{D}+2}{2},\frac{\mathcal{D}-1}{2};\frac{1}{2} \left(\mathcal{D}^2+\mathcal{D}-2\right);-\frac{1-\mathcal{D} \lambda_{min}}{\lambda_{min}}\right),
\end{align}
where $\dd$ is the dimension of the reduced density matrix. The above formula is exact in $\dd$. 
\begin{figure}[ht]
    \centering
    \includegraphics[width=  \linewidth]{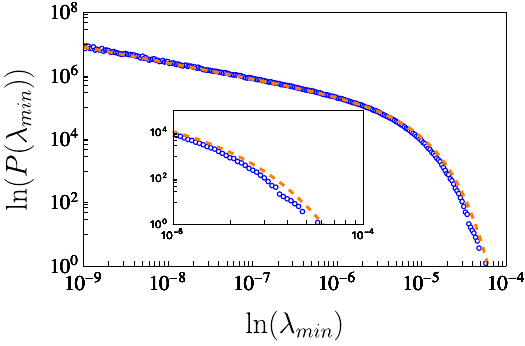}
    \caption{Distribution of minimum Schmidt eigenvalue $\lambda_{min}$ of UM matrices (blue open markers). The orange dashed line analytical result given by Eq. \eqref{eqn:lmin}. The inset shows the tail of the distribution where we observe a slight mismatch.}
    \label{fig:ummin90}
\end{figure}

In Fig. \eqref{fig:ummin90} we have plotted the numerical results obtained for the case of UM matrices ( blue open markers) with the analytical result given by Eq. \eqref{eqn:lmin}. We obtain a nice agreement between the numerical and the analytical results, suggesting that the minimum eigenvalue $\lambda_{min}$ is still strongly correlated. We however would like to note that the agreement near the tail is not good, as shown in the inset of Fig. \eqref{fig:ummin90} (contrast it with the results for Wishart matrices, shown in Appendix \ref{app:tracywidom} Fig. \eqref{fig:wlmin12}). It is ensured that the number of eigenvalues to obtain the distribution are kept same for proper comparison. The tail has better agreement for the case of Wishart distribution as can be seen in the inset of Fig. \eqref{fig:wlmin12}. These results might suggest a slightly weaker correlated smallest Schmidt eigenvalue, for the case of UM random matrices, given that Eq. \eqref{eqn:lmin} is exact in $\dd$. To further verify this we calculate first few moments, $\braket{\lambda_{min}^{k}}$, for the UM matrices and compare them with other results as well in Table \ref{tab:moments}. We observe the mismatch of the moments from the first moment itself, as compared to the value for Wishart matrices (W) which agree well for the same matrix dimensions. This implies that the agreement of numerical results with the analytical result as shown in Fig. \eqref{fig:ummin90} is not as good as it seems \footnote{With careful observation i.e. by enlarging the Fig. \eqref{fig:ummin90} it is evident that the matching is not really that well when compared to Fig. \eqref{fig:wlmin12}.}

\section{Quantum Sun model}\label{sec:qsm}

The QSM Hamiltonian is given by 
\begin{equation}\label{eqn:qsm}
    H = H_{\text{dot}} + \sum_{j=0}^{L-1} \alpha^{\mu_{j}} S^{x}_{n_{j}} S^{x}_{j} + \sum_{j=0}^{L-1} h_{j} S^{z}_{j}.
\end{equation}
The $H_{\text{dot}}$ term is modeled using a $2^{\n} \times 2^{\n}$ matrix drawn of GOE type. Formally it is given as follows 
\begin{equation}
   H_{\text{dot}} = \frac{\gamma}{\sqrt{2^{\n}+1}} \frac{1}{\sqrt{2}} (M + M^{T}),
\end{equation}
where $M$ is a matrix whose matrix elements are sampled from a standard normal distribution and the pre-factor of $\frac{\gamma}{\sqrt{2^{\n}+1}}$ ensures that at $\gamma = 1$, Hilbert-Schmidt norm $||H_{dot}||= 1$.

The second term in Eq. \eqref{eqn:qsm} described the interaction between the spin $1/2$ particles lying inside and outside of the dot. The coupling strength $\alpha^{\mu}$ is tuned using $\alpha$ and (random) distance between the coupled particles $u_{j}$. $u_{j}$ is sampled from a uniform distribution $u_{j} \in [j-\epsilon_j, j+\epsilon_j] $ with $\epsilon_j = 0.2$ for all $j$, except $\epsilon_0= 0$. 

The third term in Eq. \eqref{eqn:qsm} described the fields that acts on the spin $1/2$ particles lying outside the dot. The strength of the field $h_{j}$ is chosen from a uniform distribution, $h_{j} \in [\mathcal{S}-\delta_{\mathcal{S}}, \mathcal{\mathcal{S}} + \delta_{\mathcal{S}}]$ with $\mathcal{S}= 1$ and $\delta_{\mathcal{S}}= 0.5$. We remark that these are the standard parameters to study the model. 

The UM matrices and the QSM are conjectured to have nearly identically behaviour in the thermodynamic limit for a broad range of model parameters \cite{PhysRevResearch.6.023030}. In the following we discuss the results for the case of QSM in connection to this conjecture to further verify its validity. 

For the numerical purposes and to reduce the finite size effects, as has been found in \cite{PhysRevResearch.6.023030}, we take $\n=5$ for the QSM. We also provide the results for $\n= 3$ in appendix \ref{appendix:qsm3}, without much qualitative differences.  Further, we present our studied for $\n + L= 12$. For the numerical analysis we choose the parameter $\alpha= 0.9$, for which various other studies reveals that the eigenstates are ergodic \cite{PhysRevResearch.6.023030} in nature. However, we believe the results to hold for higher sizes as well apart from a few differences when compared with the UM random matrices. These differences might be attributed to the absence of certain interaction terms in the QSM when compared to the UM matrix. For the case of QSM we only have two-body interactions between the particles which lie within the dot (which is randomly chosen) and outside the dot. On the other hand for the case of UM random matrices a  particle $p$ outside the dot is effectively coupled to all the particles inside the block as well to all the particles $p'<p$ outside the dot as well. This is visible in the structure of the two matrices as well \cite{PhysRevResearch.6.023030} as well where for the QSM we have a sparse structure of the Hamiltonian while for the case of UM matrices we have dense structure with some fine structure, similar to the QSM, within it. 

\subsection{Eigenvector distribution}

\begin{figure}[ht]
    \centering
    \includegraphics[width=  \linewidth]{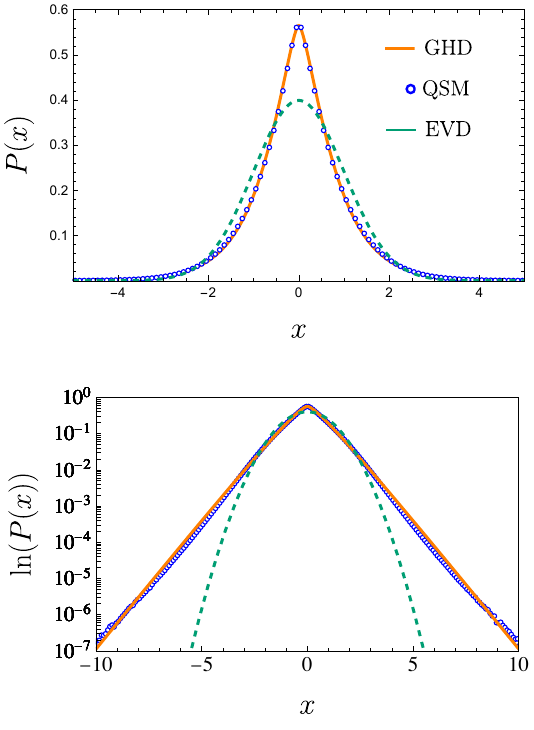}
    \caption{(Top) Distribution of rescaled eigenvector components of QSM with $\n= 5$ and $\alpha= 0.90$. Solid red line are the fitted GHD distribution given by Eq.\eqref{eq:ghd} and \textcolor{teal}{teal} dashed lines denote the standard normal distribution. (Bottom) Same as the top plot but with logarithmic $y-$ axis.}
    \label{fig:sunghdfit}
\end{figure}

The distribution of eigenvector component of the QSM are found to have similar features to that of the UM random matrices in the ergodic regime. As shown in Fig. \eqref{fig:sunghdfit}, the fitting with GHD as given by Eq. \eqref{eq:ghd}, is not satisfactory unlike the case of UM matrices and the deviations are easily detected in a logarithmic scale as shown in Fig. \eqref{fig:sunghdfit} (bottom). The fitting parameters are found to be: 
: 
\begin{center}
    $ \lambda= 1.41909 , \quad \xi= 2.66674 \times 10^{-3}$.
\end{center}

Though the plot in the linear scale might hint that the fitting is optimal; however the plot in the logarithmic $y-$axis and a careful examination of the relative error reveals that the above fitting parameters are not really optimal when compared with the case of UM random matrices. More careful analysis reveal that it might not be possible to fit satisfactorily over all the values ( both small and large values of $P(x)$). Such deviations are an example of the dissimilarities between the QSM and the UM random matrices. 

Thus the generalized hyperbolic distribution does not capture the behaviour of the distribution of eigenvectors of the QSM correctly .

\subsection{Deviations from Marchenko-Pastur distribution}

\begin{figure}[ht]
    \centering
    \includegraphics[width=  \linewidth]{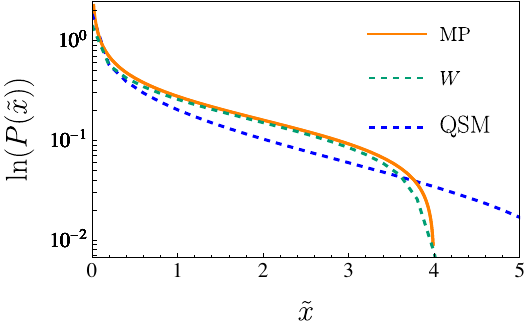}
    \caption{Distribution of rescaled eigenvalues, $\tilde{x}$ of the QSM shown as dashed blue curve. The orange solid line is the Marchenko-Pastur distribution, Eq. \eqref{eqn:mplaw} and the dashed \textcolor{teal}{teal} line is the numerical curve obtained for the trace restricted Wishart matrices of size $2^{6} \times 2^{6}$, so as to match the size of reduced density matrix for the UM matrix case. We see a clear deviation from the Marchenko-Pastur law for the case of UM matrices, specially near the tail, while the Wishart matrices match quite well with the MP law for the same size matrices.}
    \label{fig:sunmplaw}
\end{figure}

\begin{figure}[htbp]
    \centering
    \includegraphics[width=  \linewidth]{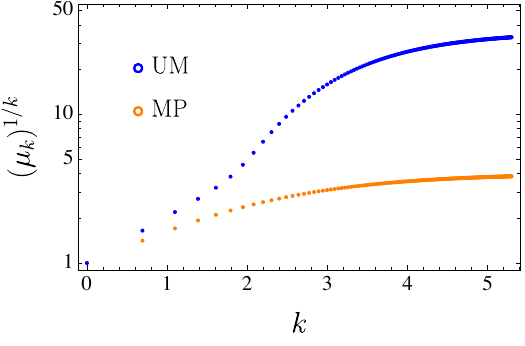}
    \caption{Comparison of moments, $(\mu_{k})^{1/k}$; $\mu_{k}$ is the $k$-th momen, of the Schmidt eigenvalues of QSM (blue markers) with the moments of Marchenko-Pastur distribution (orange markers).}
    \label{fig:sunmom12}
\end{figure}

We now consider the distribution of the Schmidt eigenvalues corresponding to the ergodic eigenstates of the QSM. The corresponding results are shown in Fig. \eqref{fig:sunmplaw}. Similar to the case of UM matrices, section \eqref{sec:ummp}, we find deviations from the MP distribution. The deviations are significant for the present case than the case of UM random matrices. To further validate this deviation from the Marchenko-Pastur law we compare $\mu^{1/k}_{k}$, where $\mu_{k}$ are the moments, of the MP distribution and the Schmidt eigenvalues of the QSM. The results are shown in Fig. \eqref{fig:sunmom12} and show significant deviations even for smaller values of $k$.

\subsection{Extreme value statistics of Schmidt eigenvalues}
\begin{figure}[ht]
    \centering
    \includegraphics[width=  \linewidth]{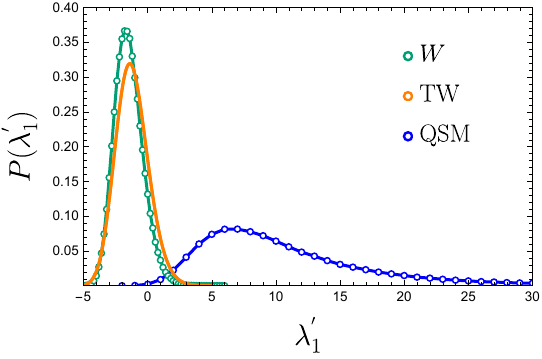}
    \caption{Distribution of maximum Schmidt eigenvalue, $\lambda'$, of the QSM ( blue markers). The results are compared to the Tracy-Widom distribution ( blue solid line). The \textcolor{teal}{teal} markers are the results for the Wishart matrices of same dimension, $2^{6} \times 2^{6}$, for the comparison purposes. The blue and \textcolor{teal}{teal} line are guide to the eye.}
    \label{fig:sun90tw}
\end{figure}

\begin{figure}[ht]
    \centering
    \includegraphics[width=  \linewidth]{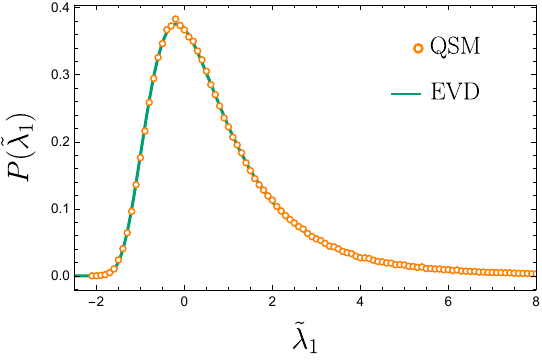}
    \caption{Distribution of rescaled eigenvalues, $\tilde{\lambda}$. The \textcolor{teal}{teal} solid line is the fitted Extreme value distribution and the orange markers are the numerical results obtained for the QSM with $\n= 3$.}
    \label{fig:sun90evd}
\end{figure}

Next we now consider the extreme, both maximum and minimum, eigenvalues of the QSM model. As we previously found for the case of UM, section \ref{sec:extum}, the maximum Schmidt eigenvalues ( suitably centered and rescaled) for the QSM also show deviations from the Tracy-Widom distribution as shown in Fig. \eqref{fig:sun90tw}. Quantitatively, the deviations are larger as compared to UM random matrices as we see the center shifted further away and peak diminished if we compare Fig. \eqref{fig:sun90tw} with Fig. \eqref{fig:umtw}.
In Fig. \eqref{fig:sun90evd}, we show that the maximum Schmidt eigenvalue  distribution ( suitably centered and rescaled) instead show agreement with the extreme value distribution. The fitting parameters are follows
\begin{center}
    $\tilde{\alpha}= 8.15\times 10^{-2}, \quad \tilde{\beta}= 1.13\times 10^{-2}$, \quad $\tilde{\xi}= -2.06\times 10^{-1}$,
\end{center}
and suggest \emph{Fr\'echet} distribution unlike the UM case where it was Weibull like. This behaviour still indicates that the maximum eigenvalue is not strongly correlation and instead is weakly correlated ( or uncorrelated) thus showing the significant deviations from the Tracy-Widom distribution.   

\begin{figure}[ht]
    \centering
    \includegraphics[width=  \linewidth]{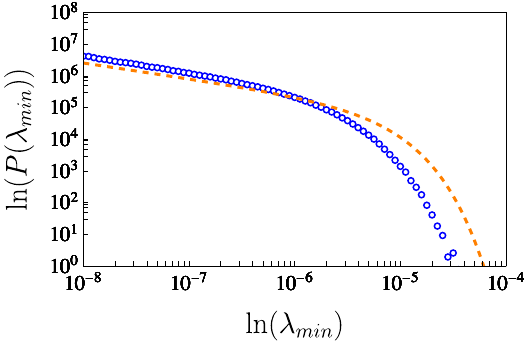}
    \caption{Distribution of minimum Schmidt eigenvalue $\lambda_{min}$ of the QSM(blue open markers) with $\n= 5$. The orange dashed line analytical result given by Eq. \eqref{eqn:lmin}.}
    \label{fig:lmin12sfinal}
\end{figure}

Finally, we compare the results of the minimum eigenvalue distribution in Fig. \eqref{fig:lmin12sfinal}. It is evident that the minimum eigenvalue shows significant deviation. To further validate this deviation we calculate the first 5 moments of the minimum eigenvalue distribution and shown them in Table \eqref{tab:moments}. The moments of the minimum Schmidt eigenvalue distribution, $\lambda_{min}$ can be analytically obtained \cite{majumdar2008exact} and given as
\begin{align} \label{eq:momlmin}
    \mu_{k}=  \frac{\left(\Gamma (k+2) \Gamma \left(k+\frac{1}{2}\right) \Gamma (\dd+1) \Gamma \left(\frac{\dd^2}{2}\right)\right)}{2^{\dd-1} \Gamma \left(\frac{\dd}{2}\right) \Gamma \left(\frac{\dd^2}{2} + k\right) \Gamma \left(\frac{\dd+3}{2} + k\right)} \\ \nonumber 
    \times \, _2F_1\left(k+2, k+\frac{1}{2}; \frac{\dd+3}{2} + k; 1-\dd\right).
\end{align}

We notice that even for the smallest moment $\mu_{1}$ there are deviations for the case of UM matrices and QSM while for the trace restricted Wishart matrices the values match with the theoretical value ( upto 3 significant digits).
\begin{table*}[htbp]
    \centering
    \begin{tabular}{|l|l|l|l|l|} \hline
         $k$& Analytical (Eq. \eqref{eq:momlmin})   & W & UM &  QSM \\ \hline
        1 & 2.64752$\times 10^{-6}$   & $2.64608 \times 10^{-6}$  & $2.32772\times10^{-6}$&$1.04794 \times 10^{-6}$\\ \hline
        
        2 & 2.22945 $\times 10^{-11}$ & $2.22818\times10^{-11}$   & $1.73073\times10^{-11}$ & $3.95408 \times 10^{-12}$ \\ \hline
        
        3 & $3.23457 \times 10^{-16}$  & $3.24205\times10^{-16}$  & $2.22347\times10^{-16}$&  $2.83556 \times 10^{-17}$\\ \hline
        
        4 & $6.68186\times10^{-21}$  & $6.76483\times10^{-21}$ & $4.07145\times10^{-21}$&  $3.16013 \times 10^{-22}$\\ \hline
        
        5 & $1.77884\times10^{-25}$  &$1.84853\times10^{-25}$  &$9.62221\times10^{-26}$ & $4.93093 \times 10^{-27}$\\ \hline
    \end{tabular}
    \caption{The first 5 moments, $\mu_{m}$ of the minimum eigenvalue $\lambda_{min}$. The values are shown using the exact analytical result Eq. \eqref{eq:momlmin}, Wishart matrices, UM matrices and QSM respectively. We observe that apart from Wishart matrices the other two cases i.e. UM  matrices and QSM show deviation even for the smallest moment. }
    \label{tab:moments}
\end{table*}

\section{Summary and Outlook} \label{sec:summary}

This work establishes the use of extreme value statistics, both maximum and minimum, of the Schmidt eigenvalues of the reduced density matrices corresponding to the ergodic eigenstates of UM random matrices and the Quantum Sun Model (QSM). The Schmidt eigenvalues are expected to follow the joint probability density function of trace-restricted Wishart random matrices, with their limiting density given by the universal Marchenko-Pastur law \cite{mehta2004random}. However, our findings reveal deviations from this law, indicating that the eigenstates are not fully random or ergodic.

The previous studies establishes analytically and numerically that the localized-to-ergodic transition in these models occur at the precise value of $\alpha_{c}= 1/\sqrt{2}$. We find that the density of Schmidt eigenvalues does not follow the \emph{universal} Marchenko Pastur law, indicating that the eigenstates are still not ergodic (and fully random). We further find that the maximum Schmidt eigenvalue density for these models, in already established ergodic regime, still fails to follow the expected Tracy-Widom distribution. These distributions are universal limiting distributions for the case when the eigenvalues are strongly correlated, hence such a deviation from Tracy-Widom implies that the correlation among the Schmidt eigenvalues might be weaker and might follow Fisher–Tippett–Gnedenko theorem. We test this hypothesis by examining the distribution of the suitably centered and rescaled eigenvalues and find that they fit quite well with the expected extreme value distribution thus implying that the correlations are not Wishart like and the ergodic eigenstates are not \emph{truly random}. Remarkably for the trace restricted Wishart matrices minimum eigenvalues are also known exactly \cite{majumdar2008exact} and we compared the result for the UM and QSM against them. While deviations from the analytical results (which are exact in the dimension of the matrices) were observed for both the cases, UM random matrices still follow the distribution closely as compared to the QSM. 

Overall the work establishes that even in the ergodic regime, $\alpha> \alpha_{c}= 1/\sqrt{2}$, the eigenstates are not truly random. Simple breaking down of RMT-like spectral statistics isn't enough to explain this phenomenon. Extreme value statistics could provide a more sensitive and stringent test for probing the randomness of the eigenstates \cite{pal2020}. One possible reason for this is that extreme value statistics encompasses the knowledge of the probability density (the way it decays and whether it is bounded or unbounded) of the random variables under consideration as well as of the strength of the correlations, while correlation measures such as the nearest spacing ratio; $r-$value captures only the strength of short range correlations. Thus we believe that the study of higher order correlations measures thus becomes important in these cases \cite{PhysRevB.98.104305,Srivastava_2018,Buijsman_2019, xu2019,Rao_2022} to capture the strength of correlations. However, it still remains an open question to understand and determine the precise transition in the \emph{extreme value statistics}, which allow us to obtain the critical value of $\alpha$ such that the states are ergodic as well as truly random. It is also important to study the results presented here in connection to the recent result on fading ergodicity regime, which establishes a link between non-ergodic behaviour and conventional ETH \cite{PhysRevB.110.134206, świętek2024s}. It is suggested that the breakdown of conventional ETH should not be associated with the breakdown of RMT like spectral statistics as the conventional ETH breaking occurs in the regime when RMT like spectral statistics is still valid.  Finally, detailed exploration of the observed deviations from the Marchenko-Pastur law in UM random matrices and the QSM is crucial, as these systems appear to lie outside the regime where this universal law applies. Notably, similar deviations have been observed in disordered Heisenberg systems hosting many-body localized phases \cite{pal2020}. By studying these deviations alongside the results presented here, we can contribute to a deeper understanding of ergodicity, randomness, and the mechanisms underlying MBL.

\emph{Acknowledgements:}
Numerical computations were performed using the computational facilities of the Yukawa Institute for Theoretical Physics. The Yukawa Research Fellowship of the author is supported by the Yukawa Memorial Foundation and JST CREST (Grant No. JPMJCR19T2).

\appendix\label{appendix}

\section{ Extreme eigenvalue statistics of Wishart matrices}\label{app:tracywidom}

\subsection{Maximum Schmidt eigenvalue distribution}
\begin{figure}[htbp]
    \centering
    \includegraphics[width= \linewidth]{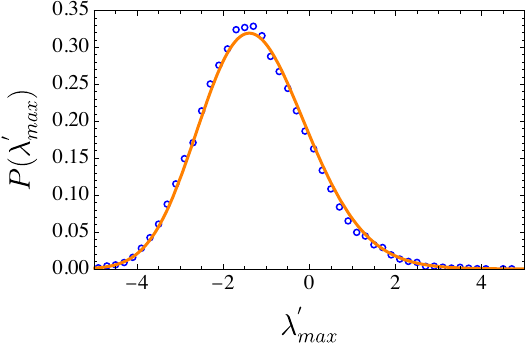}
    \caption{Distribution of $\tilde{\lambda}_{max}$ for the trace restricted Wishart matrices of size $2^{12} \times 2^{12}$. The solid red line is the Tracy-Widom distribution $F_{1}$ and blue open markers are the numerically obtained values for the Wishart matrices. The averaging is done over 30000 Hamiltonian realization.}
    \label{fig:tww12}
\end{figure}

We first provide the result for the maximum eigenvalue distribution. As was mentioned in section \ref{sec:extum} that the Tracy-Widom distribution are are limiting distribution valid in the large dimension limit of matrices. Just to support this we provide result for $2^{12} \times 2^{12}$ dimensional matrices, averaged over 30000 Hamiltonian realizations in Fig. \eqref{fig:tww12}. For the numerical purposes it suffices to obtain only the \emph{largest} eigenvalue for each of the realization using Arnoldi, thus making computation efficient. We observe that the numerical results match better as compared to Fig. \eqref{fig:umtw} for $2^{6} \times 2^{6}$ dimensional matrices.

\subsection{Minimum Schmidt eigenvalue distribution}

The probability distribution for minimum eigenvalue, $\lambda_{min}$ of trace constrained Wishart matrices is given as follows \cite{majumdar2008exact}
\begin{align}\label{eq:lmin}
    \mu_{k}=  \frac{\left(\Gamma (k+2) \Gamma \left(k+\frac{1}{2}\right) \Gamma (\dd+1) \Gamma \left(\frac{\dd^2}{2}\right)\right)}{2^{\dd-1} \Gamma \left(\frac{\dd}{2}\right) \Gamma \left(\frac{\dd^2}{2} + k\right) \Gamma \left(\frac{\dd+3}{2} + k\right)} \\ \nonumber 
    \times \, _2F_1\left(k+2, k+\frac{1}{2}; \frac{\dd+3}{2} + k; 1-\dd\right).
\end{align}
where $\mathcal{D}$ is the dimension of the Wishart matrix. 

\begin{figure}[htbp]
    \centering
    \includegraphics[width= \linewidth]{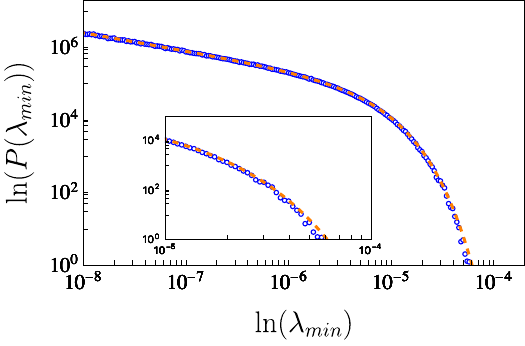}
    \caption{Distribution of $\lambda_{min}$ for the trace restricted Wishart matrices of size $2^{6} \times 2^{6}$. The solid orange line is the analytical result, Eq. \eqref{eq:lmin}  and blue open markers are the numerically obtained values for the Wishart matrices. The averaging is done over $6 \times 10^{5}$ realization. The inset shows the zoom in view of the tail.}
    \label{fig:wlmin12}
\end{figure}

In Fig. \eqref{fig:wlmin12} we show the numerical results for the $2^{6} \times 2^{6}$ dimensional matrices, with $6 \times 10^{5}$ realizations. We only obtain the smallest eigenvalue for each realization using the Arnoldi algorithm which suffices for the purpose. The match is better (which is obviously expected) with the analytical results. Also note the tail of the distribution where a better match is obtained as compared to the UM random matrices, shown in inset of Fig. \eqref{fig:ummin90}. 

\section{Quantum sun model with $\n= 3$}\label{appendix:qsm3}
We now provide the results for the QSM with $\n= 3$. It is to be noted that the results here complement the results presented in the main text and are qualitatively the same.

\subsection{Eigenvector distribution}
\begin{figure}[ht]
    \centering
    \includegraphics[width=  \linewidth]{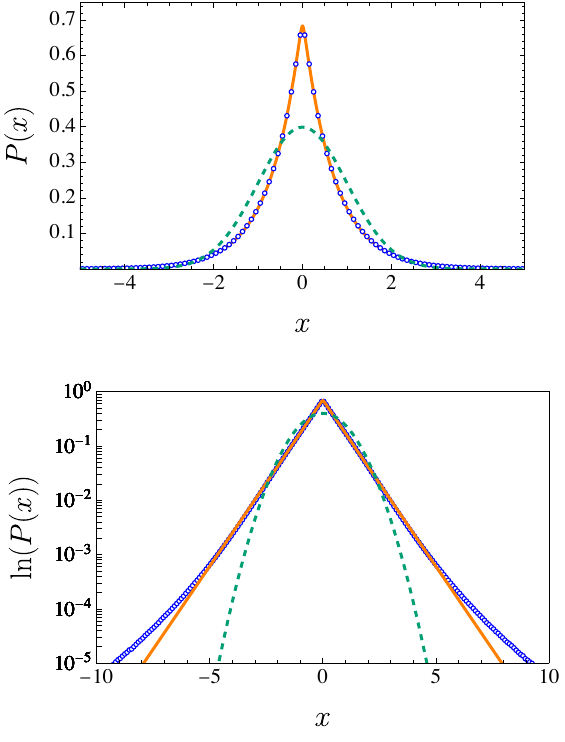}
    \caption{(Top) Distribution of rescaled eigenvector components of QSM with $\n= 3$ and $\alpha= 0.90$. Solid red line are the fitted GHD distribution given by Eq.\eqref{eq:ghd} and \textcolor{teal}{teal} dashed lines denote the standard normal distribution. (Bottom) Same as the left plot but with logarithmic $y-$ axis.}
    \label{fig:sunghdfit3}
\end{figure}

The distribution of eigenvector component of the QSM are found to have similar features to that of the UM random matrices in the ergodic regime. As shown in Fig. \eqref{fig:sunghdfit3}, they behave like GHD as given by Eq. \eqref{eq:ghd}. The fitting parameters are found to be: 

\begin{center}
    $ \lambda= 9.8352\times 10^{-1}, \quad \xi= 5.1226\times 10^{-2}$.
\end{center}
However, we obtain clear deviation from the GHD fitting at small values of $P(x)$.

\subsection{Deviations from Marchenko-Pastur distribution}

\begin{figure}[htbp]
\centering
\begin{minipage}{.45\textwidth}
  \centering
  \includegraphics[width=\linewidth]{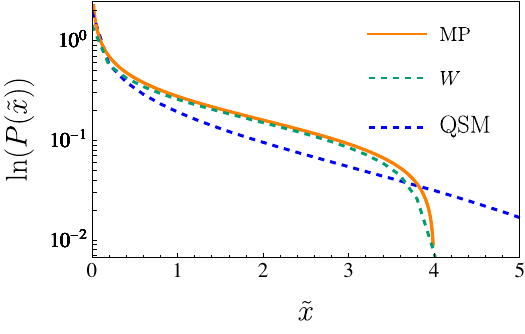}
\end{minipage}%
\hspace{.2cm}
\begin{minipage}{.45\textwidth}
  \centering
  \includegraphics[width=\linewidth]{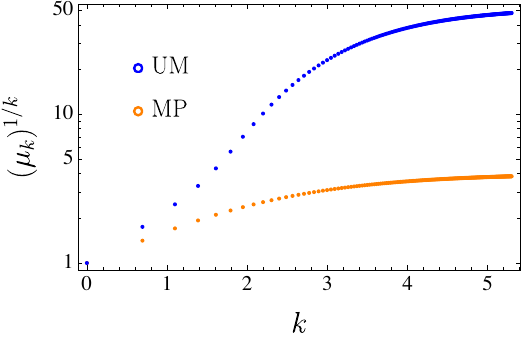}
\end{minipage}
\caption{(Top) Distribution of rescaled eigenvalues, $\tilde{x}$ of the QSM, with $\n=3$, shown as dashed blue curve. The orange solid line is the Marchenko-Pastur distribution, Eq. \eqref{eqn:mplaw} and the dashed \textcolor{teal}{teal} line is the numerical curve obtained for the trace restricted Wishart matrices of size $2^{6} \times 2^{6}$, so as to match the size of reduced density matrix for the UM matrix case. We see a clear deviation from the Marchenko-Pastur law for the case of UM matrices, specially near the tail, while the Wishart matrices match quite well with the MP law for the same size matrices. (Bottom) Comparison of moments, $(\mu_{k})^{1/k}$; $\mu_{k}$ is the $k$-th momen, of the Schmidt eigenvalues of $\n=3$ QSM (blue markers) with the moments of Marchenko-Pastur distribution (orange markers).}\label{fig:sunmplaw3}
\end{figure}

We now consider the distribution of the Schmidt eigenvalues corresponding to the ergodic eigenstates of the QSM. The corresponding results are shown in Fig. \eqref{fig:sunmplaw3}. Similar to the case of UM matrices, section \eqref{sec:ummp}, we find deviations from the MP distribution. To further validate this deviation we compare $\mu^{1/k}_{k}$, where $\mu_{k}$ are the moments, of the MP distribution and the Schmidt eigenvalues of the QSM. The results are shown in Fig. \eqref{fig:sunmplaw3} and show significant deviations even for smaller values of $k$.

\subsection{Extreme value statistics of Schmidt eigenvalues}
\begin{figure}[htbp]
\centering
\begin{minipage}{.45\textwidth}
  \centering
  \includegraphics[width=\linewidth]{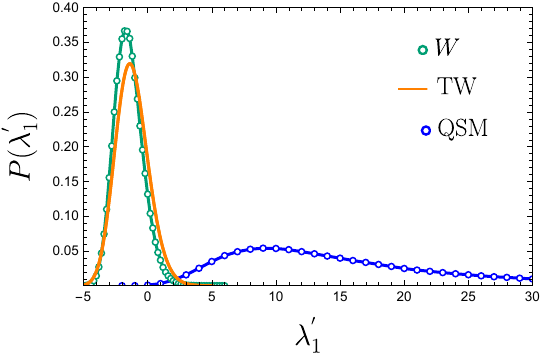}
\end{minipage}%
\hspace{.2cm}
\begin{minipage}{.45\textwidth}
  \centering
  \includegraphics[width=\linewidth]{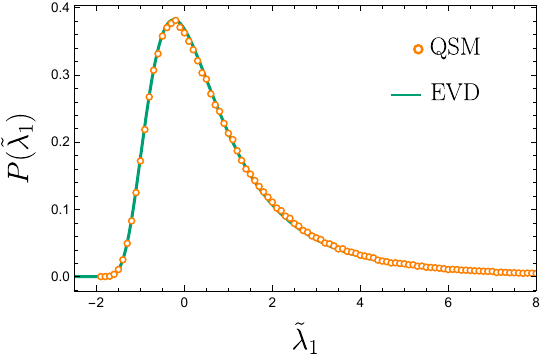}
\end{minipage}
\caption{(Top) Distribution of maximum Schmidt eigenvalue, $\lambda'$, of the QSM ( blue markers). The results are compared to the Tracy-Widom distribution ( blue solid line). The \textcolor{teal}{teal} markers are the results for the Wishart matrices of same dimension, $2^{6} \times 2^{6}$, for the comparison purposes. The blue and \textcolor{teal}{teal} solid lines are guide to the eye. (Bottom) Distribution of rescaled eigenvalues, $\tilde{\lambda}$. The \textcolor{teal}{teal} solid line is the fitted Extreme value distribution and the orange markers are the numerical results obtained for the QSM with $\n= 3$.}\label{fig:sun90evd3}
\end{figure}

Next we now consider the extreme, both maximum and minimum, Schmidt eigenvalues of the QSM model. As we previously found for the case of UM, section \ref{sec:extum}, distribution of maximum Schmidt eigenvalues ( suitably centered and rescaled) for the QSM also show deviations from the Tracy-Widom distribution as shown in Fig. \eqref{fig:sun90evd3} (left). Quantitatively, the deviation are larger as compared to UM random matrices as we see the center shifted further away and peak diminished if we compare Fig. \eqref{fig:sun90evd3} with Fig. \eqref{fig:umtw}.

In Fig. \eqref{fig:sun90evd}, we show that the maximum Schmidt eigenvalue  distribution ( suitably centered and rescaled) instead show agreement with the extreme value distribution. The fitting parameters are 

\begin{center}
    $\alpha= 8.98 \times 10^{-2},\quad \beta= 1.72 \times 10^{-2}$, \quad $\xi= -2.84 \times 10^{-1}$,
\end{center}

and suggest \emph{Fr\'echet} distribution unlike the UM case where it was \emph{Weibull} like. This behaviour still indicates that the maximum eigenvalue is not strongly correlated and instead is weakly correlated, thus showing the significant deviations from the Tracy-Widom distribution.

\begin{figure}[h]
    \centering
    \includegraphics[width= \linewidth]{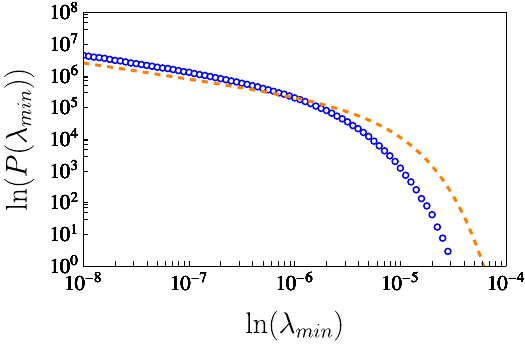}
    \caption{Distribution of minimum Schmidt eigenvalue $\lambda_{min}$ of the QSM(blue open markers) with $\n= 3$. The orange dashed line analytical result given by Eq. \eqref{eqn:lmin}.}
    \label{fig:lmin12sfinal3}
\end{figure}

Finally, we compare the results of the minimum eigenvalue distribution in Fig. \eqref{fig:lmin12sfinal3}. It is evident that the minimum eigenvalue shows a significant deviation. 

\bibliography{references}

\end{document}